\documentclass[12pt]{iopart}
\usepackage{cite}
\usepackage{color,soul}
\usepackage{graphicx}

  \expandafter\let\csname equation*\endcsname\relax
  \expandafter\let\csname endequation*\endcsname\relax

\usepackage{amsmath}

\begin{document}
\title{Effects of nitridation on SiC/SiO$_2$ structures studied by hard X-ray photoelectron spectroscopy}

\author{Judith Berens$^1$, Sebastian Bichelmaier$^1$, Nathalie K. Fernando$^2$, Pardeep K. Thakur$^3$, Tien-Lin Lee$^3$, Manfred Mascheck$^4$, Tomas Wiell$^5$, Susanna K. Eriksson$^5$, J. Matthias Kahk$^6$, Johannes Lischner$^6$, Manesh V. Mistry$^7$, Thomas Aichinger$^8$, Gregor Pobegen$^1$ and Anna Regoutz$^{2,*}$}

\address{$^1$ Kompetenzzentrum f\"{u}r Automobil- und Industrieelektronik GmbH, Europastra{\ss}e 8, 9524 Villach-St. Magdalen, Austria.}
\address{$^2$ Department of Chemistry, University College London, 20 Gordon Street, London, WC1H~0AJ, United Kingdom.}
\address{$^3$ Diamond Light Source, Harwell Science and Innovation Campus, Didcot, OX11 0DE, United Kingdom.}
\address{$^4$ Scienta Omicron GmbH, Limburger Strasse 75, 65232 Taunusstein, Germany.}
\address{$^5$ Scienta Omicron AB, P.O. Box 15120, 750 15 Uppsala, Sweden.}
\address{$^6$ Department of Materials, Imperial College London, South Kensington, London SW7 2AZ, United Kingdom and the Thomas Young Centre for Theory and Simulation of Materials.}
\address{$^7$ Department of Physics and Astronomy, University College London, Gower Street, London WC1E 6BT, United Kingdom.}
\address{$^8$ Infineon Technologies Austria AG, Siemenstra{\ss}e 2, 9500 Villach, Austria.}
\ead{a.regoutz@ucl.ac.uk}

\begin{abstract}
SiC is set to enable a new era in power electronics impacting a wide range of energy technologies, from electric vehicles to renewable energy. Its physical characteristics outperform silicon in many aspects, including band gap, breakdown field, and thermal conductivity. The main challenge for further development of SiC-based power semiconductor devices is the quality of the interface between SiC and its native dielectric SiO$_2$. High temperature nitridation processes can improve the interface quality and ultimately the device performance immensely, but the underlying chemical processes are still poorly understood. Here, we present an energy-dependent hard X-ray photoelectron spectroscopy (HAXPES) study probing non-destructively SiC and SiO\textsubscript{2} and their interface in device stacks treated in varying atmospheres. We successfully combine laboratory- and synchrotron-based HAXPES to provide unique insights into the chemistry of interface defects and their passivation through nitridation processes. 
\end{abstract}
\submitto{J. Phys. Energy}
\maketitle

\section{Introduction}

The rapid development and increasing use of electric vehicles and renewable energy is putting ever higher demands on the electronics that are at the heart of these technologies. Power electronics play a key role in controlling and converting the different forms of energy into usable electricity. They enable the delivery of electricity from the source to the end user application with maximum efficiency of transmission, distribution and consumption.\cite{2019PowerGrid} With the increasing deployment of advanced energy technologies and an overall move towards electrical energy, power electronics are crucial to enable the conversion of the energy produced by e.g. solar and wind power into electrical grid compatible forms. Beyond electric grid applications, power electronics are used in many consumer products, most importantly in electric vehicles. Traditional Si-based devices have reached the physical and material limits of what is possible, such as breakdown voltage and limited power dissipation due to thermal conductivity,\cite{Fiorenza2019CharacterizationReview,Kimoto2014FundamentalsApplications} and new materials are starting to surpass Si. Wide band gap materials, including SiC and GaN, have superior characteristics compared to Si and are increasingly taking over the main application areas of power electronics as they offer great improvements in higher power, improved thermal behaviour, and better efficiency. SiC in particular has great potential to become \textit{the} material to replace Si in many semiconductor device applications and the total market for SiC power devices is expected to exceed \$1.5 billion by 2023.\cite{Ferreira2019RoadmapEdition} SiC's material properties, which include a wide band gap, high thermal conductivity, and high breakdown field, make it the ideal semiconductor for future metal-oxide-semiconductor (MOS) devices.\cite{Weitzel1996SiliconDevices, Kimoto2015MaterialAnnealing,She2017ReviewAnd} The increasing demands on energy saving, size reduction, system integration, and improved reliability of power electronics have pushed SiC to the forefront of emerging materials. Its increased reliability, higher operating capability in both power and temperature, increased efficiency, and reduced size make it perfect for both electric vehicles and renewable energy industries. Inverters in these applications are subjected to extreme conditions, e.g. large operating temperature ranges and high power loads. Beyond its ability to address these requirements, SiC also maximises power conversion efficiency in electric vehicles resulting in an overall weight and size reduction along with increased efficiency and robust characteristics, significantly improving mileage ranges enabling overall energy savings.\\

With the immense potential for SiC to contribute to the ongoing changes in the energy landscape, intense effort focuses on further optimisation of device performance and development of ever more advanced device generations. The main obstacle for SiC to enable the usage in low-voltage classes below approximately 500 V is the low quality of its interface to its native dielectric SiO$_2$. Although SiO$_2$ can be easily grown on SiC, the interface defect densities are higher than in Si based devices leading to degradation of channel electron mobility and changes of the threshold voltage in combination with a potentially decreased reliability.\cite{Deak2007TheSiC/SiO2interface} For power electronic applications, it is the four layer hexagonal (4H) polymorph of SiC that is predominantly used. It has a band gap of 3.26 eV and defects which e.g. in silicon are within the conduction band lie within the band gap of SiC. The types of defects postulated around the SiC/SiO\textsubscript{2} interface include dangling bonds in SiC, defects in the SiO\textsubscript{2}, silicon oxycarbides (SiO\textsubscript{x}C\textsubscript{y}) and silicon oxynitrides (SiO\textsubscript{x}N\textsubscript{y}).\cite{Deak2007TheSiC/SiO2interface,Onneby1997SiliconInterface,Kobayashi2003Interfacemn0001/m,AminiMoghadam2016ActiveReview,Gruber2018ImpactInterface,Pitthan2015SiCFilms} In order to improve performance and reliability of SiC-based devices, SiC/SiO$_2$ stacks are subjected to high temperature thermal treatments in nitrogen-containing atmospheres. NO is the most widely explored annealing atmosphere and consistently shows great improvement of device performance. NH\textsubscript{3} has attracted attention as an alternative to NO as previous studies indicate that it may be able to compensate defects the NO anneal cannot passivate, in particular on the SiO\textsubscript{2} side of the interface.\cite{Pitthan2015SiCFilms,Baumvol1996ThermalHydrogen} However, devices treated under NH\textsubscript{3} show a reduction of oxide dielectric strength, which is thought to be caused by incorporation of nitrogen not just at the interface but in the bulk oxide. Overall, the nitrogen incorporated during nitridation can dramatically reduce interface defects leading to overall better device performance, however, the details of the underlying processes and how this reduction of interface defects occurs is still not well understood limiting further optimisation of the nitridation techniques\cite{Li1999InvestigationSpectroscopy,Jamet2001PhysicalSiC}. One reason for this limitation is that the characterisation of heterostructures, including buried layers and interfaces within them, presents a challenge for many established characterisation techniques and necessitates the use of very advanced techniques.\cite{Fiorenza2019CharacterizationReview,Woerle2019Two-dimensionalInterface,Isomura2019DistinguishingSpectroscopy,Gruber2018ImpactInterface,Cottom2018RecombinationCalculations,Umeda2019ElectricallyOxidation,Umeda2020CarbonInterface} Highly optimised, state-of-the-art electrical characterisation techniques can provide important information on the nature of defects in device architectures. Comparison to theoretical calculations as well as physical characterisation techniques are almost always required to unpack the underlying complex chemistry and physics. A recently very successfully employed technique to probe interface defects in SiC/SiO\textsubscript{2} is electrically detected magnetic resonance (EDMR), which can be used to probe even very small defect densities.\cite{Cottom2018RecombinationCalculations,Umeda2019ElectricallyOxidation,Umeda2020CarbonInterface} Scanning transmission electron microscopy (STEM) in combination with electron energy loss spectroscopy (EELS) is one of the most widely employed physical characterisation techniques giving structural and elemental maps of multilayer device stacks.\cite{Gruber2018ImpactInterface,Fiorenza2019CharacterizationReview} However, both electrical and microscopy techniques do not provide direct characterisation of local element-specific chemical information and it is an insurmountable challenge to identify states specific to the interface.\\

An established materials characterisation technique that promises to deliver this information and which has been applied extensively to the investigation of SiC/SiO\textsubscript{2} structures after nitridation, contributing to our current understanding of the system, is X-ray photoelectron spectroscopy (XPS).\cite{Onneby1997SiliconInterface,Li1999InvestigationSpectroscopy,Hornetz1994ARXPSSurfaces, Zhu2011ChemicalSpectroscopy, Woerle2017ElectronicARPES} We could recently show that soft X-ray photoelectron spectroscopy (SXPS) is a powerful technique to probe the chemical state of the SiC/SiO$_2$ system.\cite{Regoutz2018InterfaceAtmospheres} The sensitivity of the technique to differences in chemical environment can be used to understand how nitrogen passivates the interface defects and in turn provide information on the defects initially present. However, due to the limited information depth of soft X-rays, depth profiling using argon sputtering has to be employed to make the interface accessible for measurement. Through careful optimisation of sputtering conditions, sputtering artefacts can be minimised, but some uncertainty remains over whether interface states and the state of buried layers are fully preserved. In contrast, hard X-ray photoelectron spectroscopy (HAXPES) enables the study of such systems without the need for any sample back preparation due to the increase in probing depth when using higher X-ray energies. Whilst XPS has been used extensively to study SiC/SiO\textsubscript{2} structures, HAXPES studies are rare and have not yet been used to perform broad, systematic studies.\cite{Hamada2017AnalysisNO-POA,Yamashita2019SpectroscopicInterface}\\

Here, we present energy-dependent HAXPES results of SiC/SiO\textsubscript{2} stacks after nitridation in a variety of annealing environments, combining for the first time both laboratory- and synchrotron-based HAXPES and providing unique insights into chemical changes in both the carbide and oxide layers, as well as their interface. Four annealing atmospheres are compared, including N\textsubscript{2}, which acts as a reference, NO, NH\textsubscript{3} and a combinatorial NO + NH\textsubscript{3} process. Depth distribution functions are calculated for the different samples to model the information depth of the experiments. Changes in the nitrogen distribution after the use of different annealing atmospheres are clearly detected in the core level spectra and the chemical state of nitrogen is analysed in detail. The observed nitrogen species within the SiO\textsubscript{2} layer and at the interface enable insights into the nature of the defects passivated by nitrogen. Furthermore, the HAXPES results are compared to previous SXPS studies and commonalities and differences are discussed.\\

\section{Methods}
The 4H-SiC/SiO\textsubscript{2} samples investigated in this work were manufactured using an industrial process. n-type doped 150 mm 4H-SiC wafers with a 4$^\circ$ offset with respect to the crystalline c-axis were used. An SiO\textsubscript{2} thin film with a target thickness of 10 nm was deposited from tetraethyl orthosilicate on the Si-face of the wafers. Following this, the wafers were subjected to high temperature treatments above 1000$^\circ$C in varying nitrogen-containing atmospheres to densify the deposited oxide and optimise the interface quality, similar to processes reported in the literature.\cite{Chung2000EffectCarbide,McDonald2003CharacterizationSiO2/4H-SiC} Here, four annealing atmospheres are compared, including nitrogen (N\textsubscript{2}), nitric oxide (NO), ammonia (NH\textsubscript{3}), and a sequence of NO followed by NH\textsubscript{3}. All samples were annealed for $>$1 hour with the duration of the combinatorial anneal of NO+NH\textsubscript{3} being $>$1 hour for NO followed by a shorter anneal for NH\textsubscript{3}. Electrical characteristics and refractive index measurements for these samples were reported previously.\cite{Regoutz2018InterfaceAtmospheres} The N\textsubscript{2} sample acts as a reference as no nitrogen is incorporated into the multilayer structures using the applied processes.\\

Hard X-ray photoelectron spectroscopy (HAXPES) was performed on two different systems. Experiments at 9 keV were performed on a HAXPES Lab laboratory-based system from Scienta Omicron. This system uses a monochromated, microfocused Ga K$_\alpha$ X-ray source giving a photon energy of 9.25 keV, further referred to as 9 keV for simplicity. A Scienta Omicron EW4000 hemispherical electron energy analyser is used, with a maximum  acceptable kinetic energy of 12 keV and a large acceptance angle of $\pm$30$^\circ$. Samples were measured in grazing incidence geometry with the angle between incoming X-rays and sample surface being less than 3$^\circ$. The system has been described in detail elsewhere.\cite{Regoutz2018ASystem} Experiments at 4 and 6 keV were performed at beamline I09 at Diamond Light Source.\cite{Lee2018ASource} A double-crystal Si (111) monochromator was used to select 4 and 6 keV photons. In addition Si (022) and Si (004) channel-cut crystals were employed to achieve the final energy resolution for 4 and 6 keV, respectively. The final excitation energies were 4.062 keV and 5.922 keV, which will be further referred to as 4 and 6 keV for simplicity. Beamline I09 is equipped with a VG Scienta EW4000 electron energy analyzer with $\pm$30$^\circ$ angular  acceptance. Samples were measured in grazing incidence geometry with the angle between incoming X-rays and sample surface being less than 5$^\circ$.\\

\section{Results \& Discussion}

\subsection{Depth profile of SiC/SiO\textsubscript{2} heterostructures}

\begin{figure}
 \hfill \includegraphics[width=0.65\textwidth]{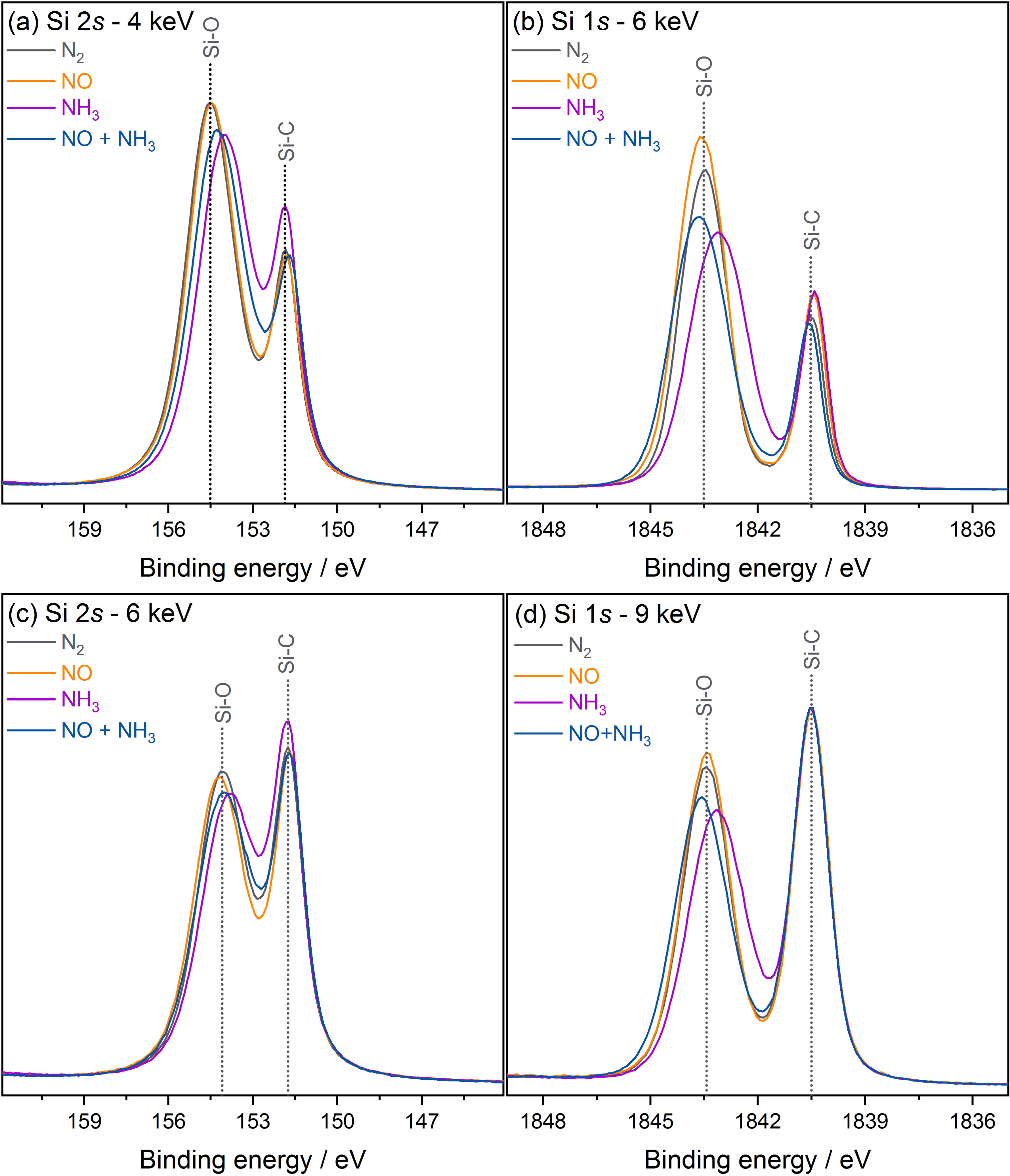}
 \caption{Si core level spectra of the four samples collected at varying X-ray excitation energies, including (a) Si 2\textit{s} at 4 keV, (b) Si 1\textit{s} and (c) Si 2\textit{s} at 6 keV, and (d) Si 1\textit{s} at 9 keV.}
  \label{fig:Si_CL_comp}
\end{figure}

\begin{figure}
 \hfill\includegraphics[width=0.84\textwidth]{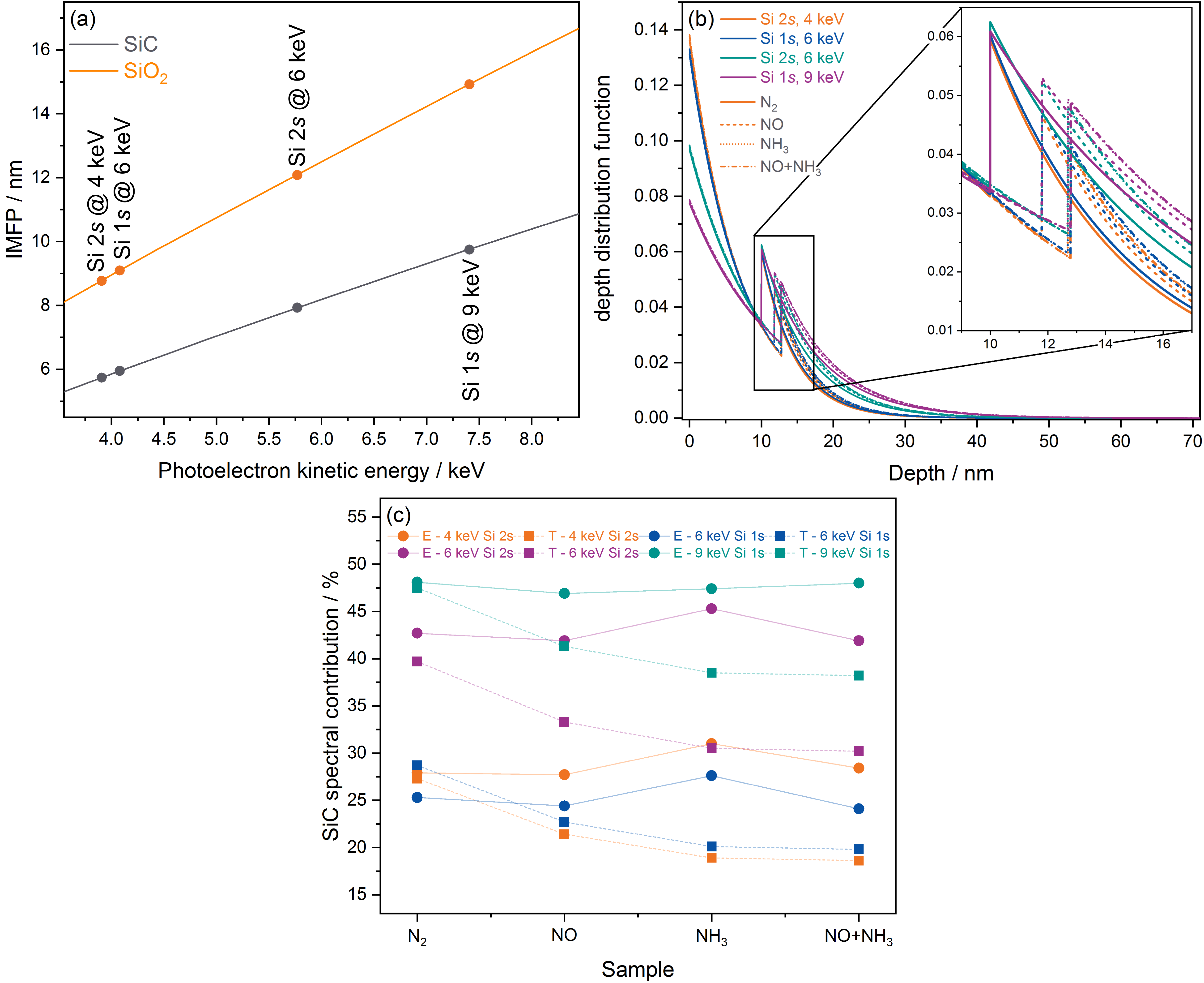}
 \caption{Probing depth of HAXPES for SiC/SiO\textsubscript{2} heterostructures. (a) Calculated inelastic mean free paths (IMFPs) for the Si core levels investigated from reference \cite{Shinotsuka2019CalculationsAlgorithm}. (b) Depth distribution functions for the Si core levels for all four samples. The inset shows an expanded view of the transition region between SiO\textsubscript{2} and SiC. (c) Comparison of SiC spectral contribution of the Si core levels for both theory (T) and experiment (E).}
  \label{fig:IMFP}
\end{figure}

In order to create a non-destructive depth profile of the SiC/SiO\textsubscript{2} heterostructures, energy-dependent HAXPES core level spectra were collected across three excitation energies of 4, 6, and 9 keV. Figure \ref{fig:Si_CL_comp} shows Si 1\textit{s} and 2\textit{s} core levels across the three excitation energies for the four samples treated in varying nitrogen atmospheres. Due to the increased depth information in HAXPES, even without the sputtering needed for SXPS, both SiO\textsubscript{2} (higher binding energy (BE) contribution) and SiC (lower BE contribution) signals are visible simultaneously at all X-ray excitation energies. Depending on the X-ray excitation energy as well as the binding energy of the core level in question the inelastic mean free path (IMFP) and therefore the probing depth changes significantly. The probing depth in XPS is defined as the average depth from which 95\% of the photoelectrons are derived. This generally equates to three times the IMFP and therefore changes in the IMFP lead to the variations in the relative Si core level ratios of the signal detected for SiO\textsubscript{2} and SiC observed in the present samples. Figure \ref{fig:IMFP}(a) shows the theoretical IMFPs for both SiO\textsubscript{2} and SiC for the four core levels investigated here. The IMFP values were taken from the work by Shinotsuka \textit{et al.} and extracted values for the kinetic energies of the Si 1\textit{s} and 2\textit{s} core levels are summarised in Table \ref{tab:IMFPs}.\cite{Shinotsuka2019CalculationsAlgorithm} The theoretical IMFPs generally overestimate the probing depth of HAXPES, and effective attenuation lengths (EAL) are shorter than predicted. In a recent paper Solokha \textit{et al.} explored this difference for silicon across a kinetic energy range from 1.5 keV to 8 keV at beamline I09.\cite{Solokha2018EffectiveMembranes} It is clear that the experimentally measured EALs are significantly smaller than the predicted EALs and IMFPs. In order to take this into account in the present work, we have corrected the predicted IMFPs to be 80\% of their original value reflecting the difference in experimental and theoretical values reported for Si by Solokha \textit{et al.} (see Table \ref{tab:IMFPs} for the values of IMFP\textsubscript{corr} used).\\

The relative intensities of the core level peaks of the overlayer (SiO$_2$) and the substrate (SiC) are determined by the depth distribution function (DDF). The DDF is defined as the probability that a photoelectron leaving the surface originated from a given depth measured normally from the surface into the material. All samples are based on bulk SiC wafers with a SiO\textsubscript{2} overlayer with varying thickness after nitridation. The SiO\textsubscript{2} thicknesses from capacitance-voltage characterisation reported in our previous paper have been used as the overlayer thicknesses, which are 10.0 nm (N\textsubscript{2}), 11.8 nm (NO), 12.7 nm (NH\textsubscript{3}), and 12.8 nm (NO+NH\textsubscript{3}), respectively.\cite{Regoutz2018InterfaceAtmospheres} A detailed description of how the DDF intensity profiles shown in Figure \ref{fig:IMFP}(b) were calculated is included in the Supplementary Information.\\

\begin{table}
\caption{\label{tab:IMFPs}Calculated and corrected Inelastic Mean Free Paths (IMFP and IMFP\textsubscript{corr}) for SiC and SiO\textsubscript{2} at the kinetic energies of the Si 1\textit{s} and 2\textit{s} core levels investigated. The calculated values are extracted from \cite{Shinotsuka2019CalculationsAlgorithm}.}
\resizebox{\textwidth}{!}{
\begin{tabular}{@{}lccccccc}
\br
Core level  & h$\nu$ / eV &   Av. BE / eV & KE / eV & IMFP (SiO\textsubscript{2}) / nm &  IMFP (SiC) / nm & IMFP\textsubscript{corr} (SiO\textsubscript{2}) / nm &  IMFP\textsubscript{corr}  (SiC) / nm\\
\mr
Si 2\textit{s} & 4062  & 153 & 3909  & 8.8 & 5.7 & 7.0 & 4.6 \\
Si 1\textit{s} & 5922  & 1842 & 4080  & 9.1 & 6.0 & 7.3 & 4.8 \\
Si 2\textit{s} & 5922  & 153 & 5769  & 12.1 & 7.9 & 9.7 & 6.3 \\
Si 1\textit{s}  & 9250  & 1842 & 7408  & 14.9 & 9.8 & 11.9 & 7.8 \\
\br
\end{tabular}}
\end{table}

From integration of the relevant sections of the DDF in Figure \ref{fig:IMFP}(b) the SiC contribution to the total signal can be calculated and compared to peak fit results of the Si core levels (see Figure \ref{fig:IMFP}(c)). This approach gives values in good agreement between theory and experiment for the N\textsubscript{2} sample across all excitation energies and core levels explored. In contrast to the N\textsubscript{2}, where nitrogen is not incorporated across the multilayer stack, the NH\textsubscript{3} and NO+NH\textsubscript{3} samples show a strong deviation between the experimentally observed signal contributions and the theoretically expected ones. This is due to large amounts of nitrogen being present in the SiO\textsubscript{2} layer after nitridation, which will be discussed further below. The incorporation of nitrogen significantly influences the IMFP, but this cannot be taken into account in the currently available methods and all results are based on pure SiO\textsubscript{2}.\\

\begin{figure}
\centering
 \hfill\includegraphics[width=0.65\textwidth]{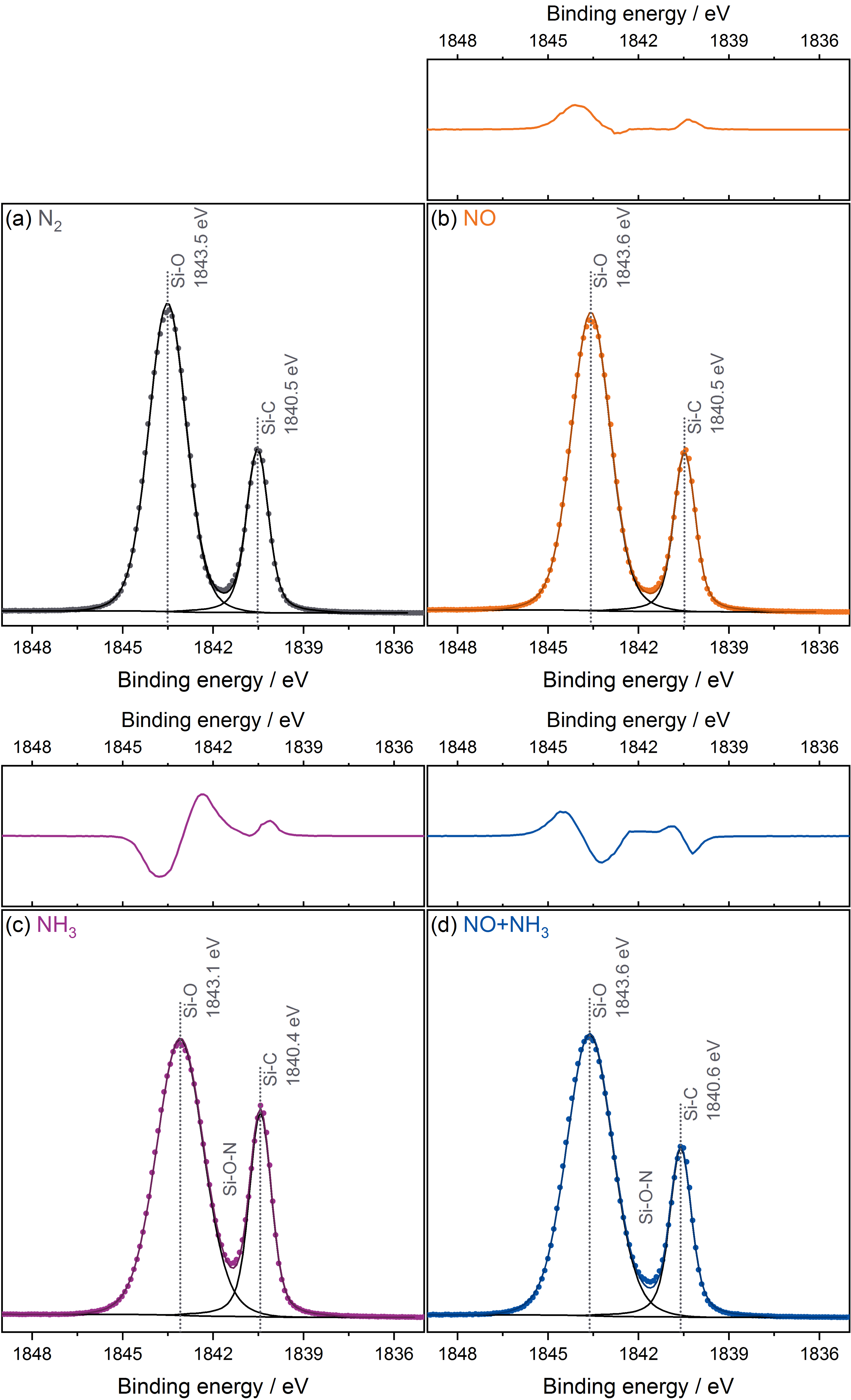}
 \caption{Peak fitted Si 1\textit{s} core level spectra of the four samples collected at h$\nu$ = 6 keV. The top graphs show the difference plots relative to the N\textsubscript{2} reference sample for the NO, NH\textsubscript{3} and NO+NH\textsubscript{3} samples.}
  \label{fig:Si1s_6keV_comp}
\end{figure}

\subsection{The effect of nitridation on SiO\textsubscript{2} and SiC}

The effect of nitridation on the SiO\textsubscript{2} layer can be observed in the line shape and BE positions of the Si core level spectra (see Figure \ref{fig:Si_CL_comp}). Whilst the bulk SiC contribution remains at a constant BE across all measurements the higher BE contribution assigned to SiO\textsubscript{2} shows significant changes in BE position as well as overall lineshape after nitridation treatments. In order to quantify these changes a peak fit analysis was conducted extracting the peak positions, full width half maxima (FWHM), and area ratios, which are summarised in Table \ref{tab:CL}. Figure \ref{fig:Si1s_6keV_comp} shows the peak fitted Si 1\textit{s} core levels collected at 6 keV, which are representative for the peak fits conducted for all core levels. In addition, difference plots relative to the N\textsubscript{2} sample are shown to aid interpretation of small spectral differences. The bulk SiC contribution to the different Si core levels remains at a constant binding energy (BE) across all nitridation atmospheres indicating that no significant changes to the bulk SiC occur, excluding the incorporation of nitrogen in the SiC substrate. The slight differences in total intensity of the SiC contribution are due to the varying SiO\textsubscript{2} overlayer thickness. In contrast to the SiC contribution, the SiO\textsubscript{2} feature at higher BE shows considerable variations in its overall energy position. In the NH\textsubscript{3} and NO+NH\textsubscript{3} treated samples, the shift of the SiO\textsubscript{2} peak is a result of the above mentioned incorporation of nitrogen species in the SiO\textsubscript{2} leading to changes of the chemical environments. This is consistent with changes in the refractive indeces of the oxide layers, which are 1.4572 (NO), 1.4684 (NH\textsubscript{3}), and 1.4644 (NO+NH\textsubscript{3}), respectively. Whilst the refractive index of the NO sample is close to standard SiO\textsubscript{2}, the values increase significantly for NH\textsubscript{3} and NO+NH\textsubscript{3} treated samples. The changes in the core spectra shown in Figure \ref{fig:Si_CL_comp} are not constant across the varying excitation energies due to the signal probing different regions of the oxide. This will be discussed in more detail in the context of the nitrogen spectra. The incorporation of nitrogen needs to be monitored closely in nitridation processes, as it can be detrimental to the overall device characteristics due to the resulting increases in oxide trap density leading to a degradation of the dielectric behaviour of SiO\textsubscript{2}.\cite{Berens2019NH3Reliability}\\

When comparing the SiO\textsubscript{2} and SiC contributions a much larger FWHM of the SiO\textsubscript{2} compared to the SiC peak is observed at all excitation energies and for both the Si 1\textit{s} and 2\textit{s} core levels. This is due to a difference in structure between the two layers. Whilst SiC is a single crystal wafer with high structural order, the SiO\textsubscript{2} film is amorphous and encompasses a range of different Si environments. This leads to a number of contributions at different BEs to the overall core level shape of the SiO\textsubscript{2} layer, which are too close in BE to be resolved individually, leading to an increase in the overall peak width observed. The NO and NO+NH\textsubscript{3} treated samples have a larger FWHM of the SiO\textsubscript{2} peak compared to N\textsubscript{2} and NO samples, again due to the incorporation of N into the SiO\textsubscript{2} layer. \\

The peak fits for the Si 1\textit{s} core level at 6 keV show two main components for SiO\textsubscript{2} and SiC in the N\textsubscript{2} and NO treated samples. Compared to all previously reported SXPS experiments, no feature below the SiC BE component could be observed. This is a direct result of the fact that sputter depth profiling is omitted when HAXPES is used, making the measurement non-destructive in nature. The lower BE features reported previously can now be unequivocally assigned to sputter artefacts in the form of partially reduced SiC. However, upon closer inspection of the peak fits for the NH\textsubscript{3} and NO+NH\textsubscript{3}, it is clear that the fitting with only two components misses some intensity intermediate in binding energy between the SiO\textsubscript{2} and SiC peaks. The difference spectra underline this mismatch further. Whilst the NO sample shows only very small deviation from the N\textsubscript{2} line shape, the NH\textsubscript{3} and NO+NH\textsubscript{3} samples show large mismatches of the signal, in particular around the SiO\textsubscript{2} contribution. This is a result of a considerable contribution from additional chemical states due to the incorporation of nitrogen, including Si-O-N and Si-C-N environments. Whilst their presence is clear in the line shapes, it is not possible to peak fit these environments reliably to distinguish them from the Si core levels due to the unknown line shape and considerable overlap with the main SiO\textsubscript{2} component.\\

\begin{table}
\caption{\label{tab:CL}Peak parameters extracted from peak fit analysis of the Si 1\textit{s} and 2\textit{s} core levels. The error of the given binding energies (BE) and full width half maxima (FWHM) is $\pm$0.1 eV.}
\resizebox{\textwidth}{!}{
\begin{tabular}{@{}llcccccc}
\br
h$\nu$ & & & SiC & & & SiO\textsubscript{2} &  \\
Core level  & Sample &   BE / eV & FWHM / eV & area / \% & BE / eV & FWHM / eV & area / \%\\

\mr
4 keV & N\textsubscript{2} &   151.9 & 1.2 & 27.9 & 154.5 & 2.1 & 72.1 \\
Si 2\textit{s} & NO &  151.8 & 1.2 & 27.7 & 154.5 & 2.1 & 72.3 \\
 & NH\textsubscript{3}  & 151.8 & 1.2 & 31.0 & 154.0 & 2.2 & 69.0 \\
 & NO+NH\textsubscript{3} & 151.7 & 1.3 & 28.4 & 154.2 & 2.2 & 71.6 \\
6 keV & N\textsubscript{2} & 1840.5 & 0.9 & 25.3 & 1843.5 & 1.5 & 74.7 \\
Si 1\textit{s} & NO & 1840.5 & 0.8 & 24.4 & 1843.6 & 1.5 & 75.6 \\
 & NH\textsubscript{3}  & 1840.4 & 0.9 & 27.6 & 1843.1 & 1.8 & 72.4 \\
 & NO+NH\textsubscript{3} & 1840.6 & 0.9 & 24.1 & 1843.6 & 1.8 & 75.9 \\
 6 keV & N\textsubscript{2} & 151.8 & 1.3 & 42.7 & 154.1 & 2.1 & 57.3 \\
Si 2\textit{s} & NO & 151.7 & 1.3 & 41.9 & 154.2 & 2.2 & 58.1 \\
 & NH\textsubscript{3}  & 151.8 & 1.3 & 45.3 & 153.9 & 2.3 & 54.7 \\
 & NO+NH\textsubscript{3} & 151.7 & 1.3 & 41.9 & 154.0 & 2.3 & 58.1 \\
9 keV & N\textsubscript{2} & 1840.5 & 1.2 & 48.1 & 1843.4 & 1.6 & 51.9 \\
Si 1\textit{s} & NO & 1840.5 & 1.2 & 46.9 & 1843.4 & 1.6 & 53.1\\
 & NH\textsubscript{3}  & 1840.5 & 1.2 & 47.4 & 1843.1 & 1.9 & 52.6\\
 & NO+NH\textsubscript{3} & 1840.5 & 1.2 & 48.0 & 1844.0 & 1.8 & 52.0 \\
\br
\end{tabular}}
\end{table}

\begin{figure}
 \hfill\includegraphics[width=0.65\textwidth]{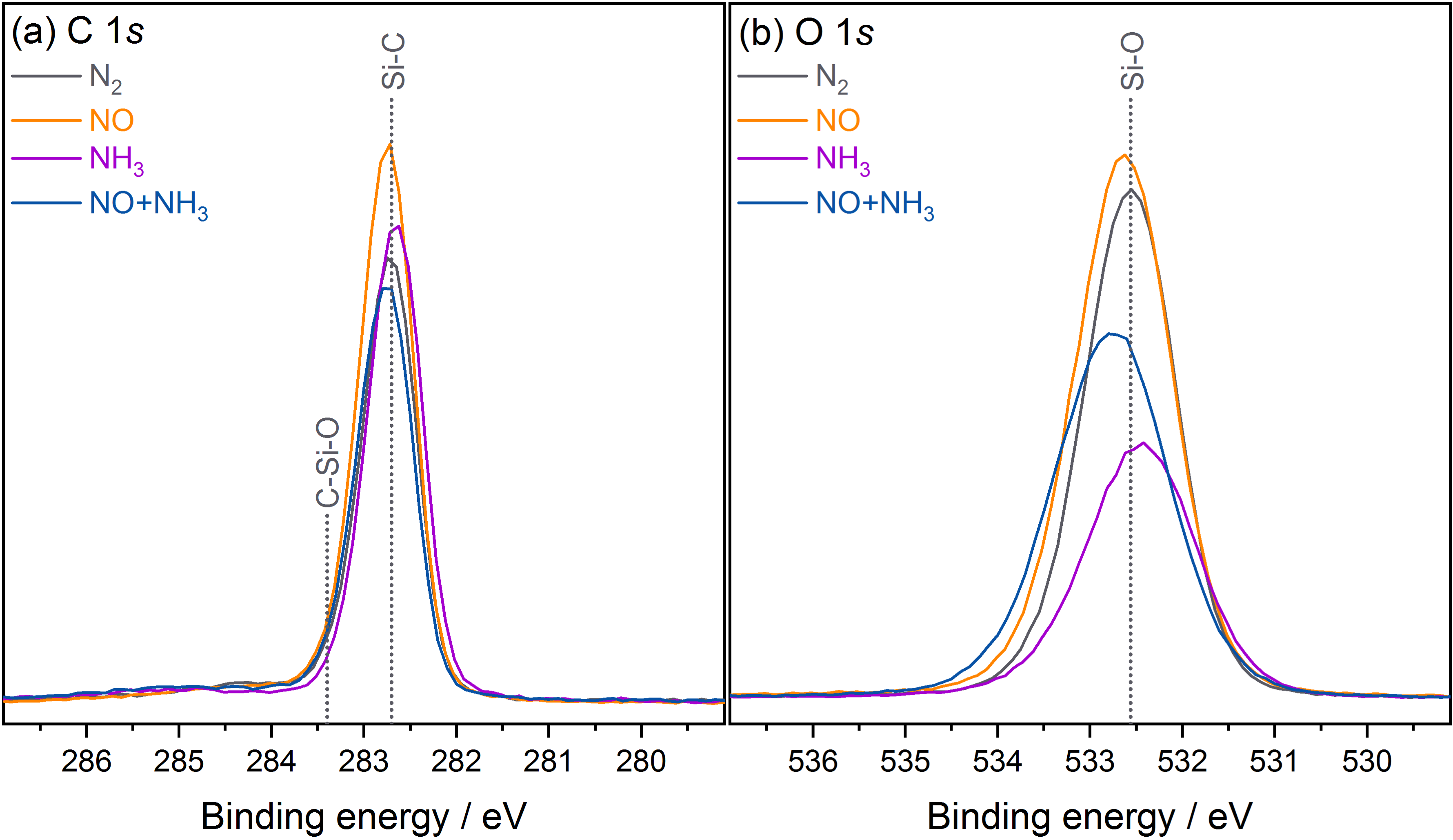}
 \caption{C and O 1\textit{s} core level spectra of the four samples collected at h$\nu$ = 6 keV.}
  \label{fig:6keV_CO_CL}
\end{figure}

In parallel to the Si core levels, the C and O 1\textit{s} core levels collected at the same excitation energies confirm the observations made from the silicon spectra and representative core levels for all four samples collected at 6 keV are shown in Figure \ref{fig:6keV_CO_CL}. The C 1\textit{s} core level remains comparable in BE position and line shape across all samples, with the main contribution at 282.7 eV BE and 0.7 eV FWHM, respectively, consistent with the observation from Si core levels that bulk SiC remains mostly unchanged after nitridation. All C 1\textit{s} core levels also exhibit a small feature towards higher BE of the main feature predominantly from C-Si-O environments in silicon oxycarbides SiO\textsubscript{x}C\textsubscript{y}, as well as some contribution from C-Si-N states. Oxycarbides are expected to contribute to the active defect population at the interface. Only small variations in relative intensities between samples are observed due to differences in signal attenuation caused by the aforementioned differences in SiO\textsubscript{2} overlayer thickness and chemistry as well as differences in the contribution from SiO\textsubscript{x}C\textsubscript{y}. In contrast, the O 1\textit{s} core level changes significantly, with the FWHM increasing from 1.2 eV for N\textsubscript{2} and NO to 1.4 eV for NH\textsubscript{3} and NO+NH\textsubscript{3} treated samples due to the incorporating of nitrogen within the SiO\textsubscript{2} layer, which will be further discussed through the analysis of the N 1\textit{s} core levels.In addition to changes in the FWHM, these samples also show a shift in their BE position for the same reasons causing a change of the chemical environments of the oxygen in parallel to the changes observed for Si. \\

\subsection{The N 1\textit{s} core level and changes across the interface}

\begin{figure}
 \hfill\includegraphics[width=0.7\textwidth]{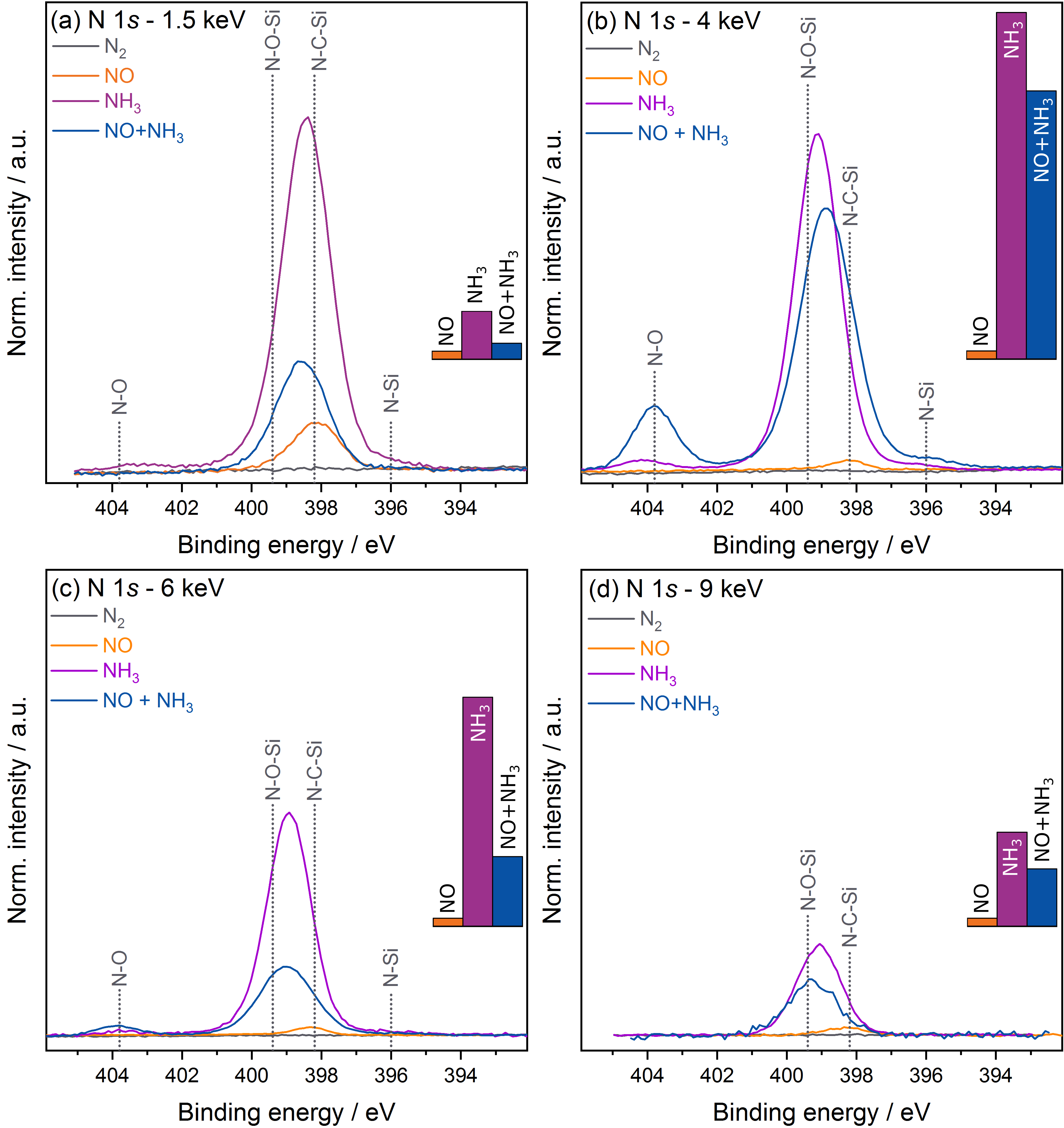}
 \caption{N 1\textit{s} core level spectra of the four samples collected at varying X-ray excitation energies, including soft X-rays at (a) 1.5 keV after sputtering, and hard X-rays at (b)-(d) 4, 6 and 9 keV. The HAXPES spectra are normalised to the SiC contribution of the C 1\textit{s} core levels of the respective samples. The insets in (a) - (d) show the relative amount of the main nitrogen species present in each sample relative to the NO signal.}
  \label{fig:N_CL_comp1_v2}
\end{figure}

The Si, C, and O core levels are useful to determine the differences in probing depth and to investigate changes in chemical environments within the SiC and SiO\textsubscript{2} layers. The N core level is of particular interest in this study to follow how it changes upon variation of the nitridation atmosphere. Figure \ref{fig:N_CL_comp1_v2} shows the N 1\textit{s} core level for the three HAXPES excitation energies as well as for the interface spectra collected after sputter depth profiling with soft X-ray photoelectron spectroscopy (SXPS), which we have discussed in detail in a previous publication.\cite{Regoutz2018InterfaceAtmospheres} One of the main advantages of HAXPES over SXPS is that buried layers and interfaces in heterostructures can be probed non-destructively, as mentioned above. Due to the necessity of sputtering to enable depth profiling with soft X-ray sources one can never completely exclude artefacts resulting from this treatment. In the case of the N 1\textit{s} spectra investigated here, the overall structure of the core levels observed in SXPS is to a large extend comparable with HAXPES. However, there are some distinct differences, which will be discussed.\\ 

\begin{figure}
 \hfill\includegraphics[width=0.65\textwidth]{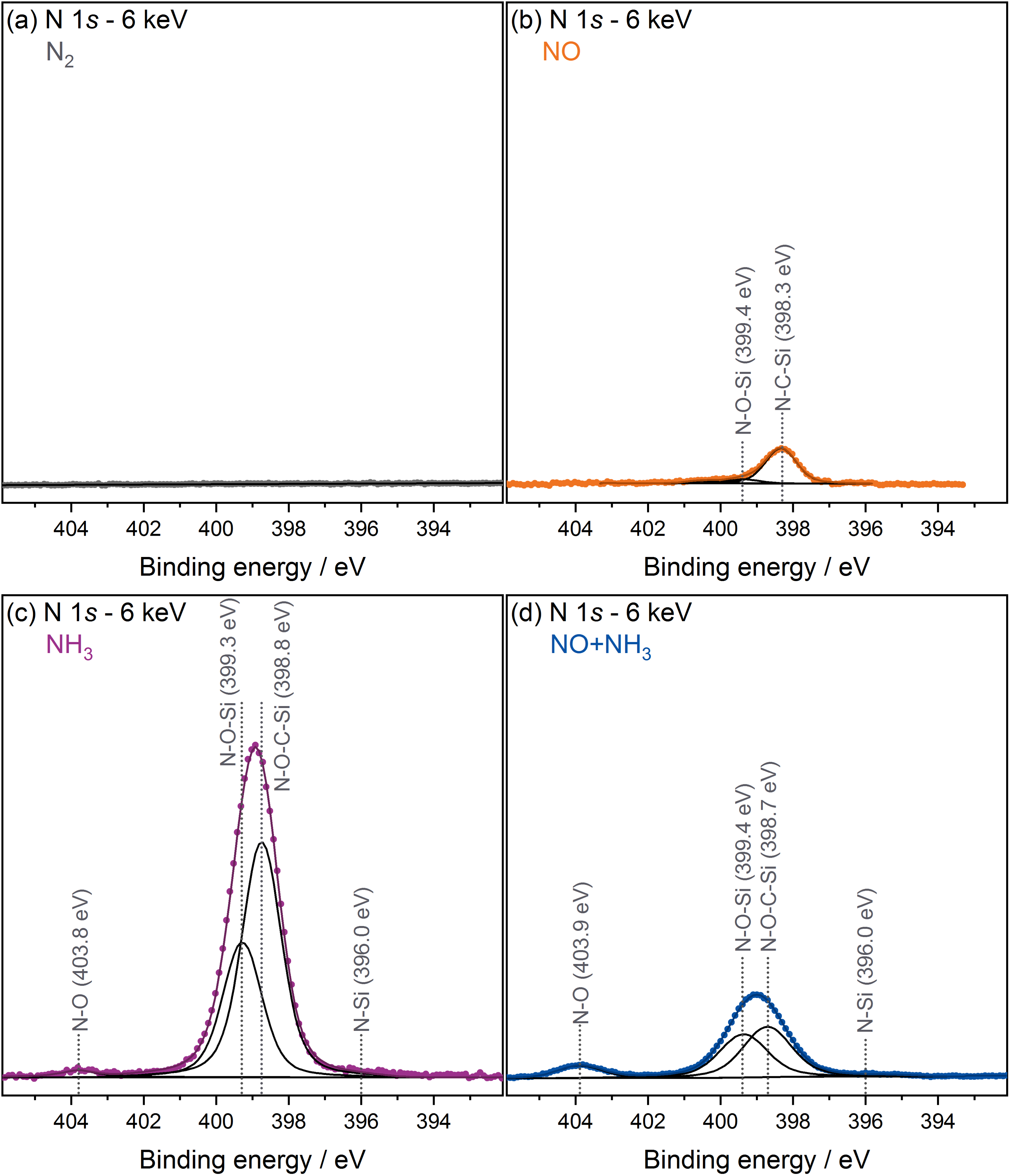}
 \caption{Peak fit analysis of the N 1\textit{s} core level spectra of the four samples collected at h$\nu$ = 6 keV.}
  \label{fig:N_CL_comp2}
\end{figure}

Comparison of the three HAXPES excitation energies reveals changes in relative intensities of the nitrogen signal. The spectra are normalised to the SiC contribution in the C 1\textit{s} core level for each excitation energy and sample allowing for comparison between the different data sets. Particularly the reduction of N signal for NH\textsubscript{3} and NO+NH\textsubscript{3} samples at higher h$\nu$ is noticeable. This is due to most of the N being incorporated at the top of the SiO\textsubscript{2} layer, but the amount of N decreasing further in the layer, which is consistent with our previous SXPS observation, which includes quantification and atomic distribution profiles. This is also the reason why the BE position of the Si 1\textit{s} and 2\textit{s} core levels is not constant across spectra collected at varying h$\nu$. The lowest BE feature in the N 1\textit{s} spectra, assigned to Si-N environments, was previously attributed to artefacts from sputter treatment in SXPS experiments. However, as the Si-N feature is observed consistently in the HAXPES spectra for the NH\textsubscript{3} and NO+NH\textsubscript{3} samples, it is intrinsic to the samples. The reduction in signal intensity of Si-N follows that of the Si-O-N feature suggesting a comparable depth distribution with the Si-N environment also predominantly being located in the SiO\textsubscript{2} layer. The N-O environments, which had very low intensity or could not be detected at all in the SXPS experiments due to destruction by sputtering, are significant particularly in the 4 keV spectra. The intensity of the N-O environment drops much faster with increasing h$\nu$ than the Si-O-N and Si-N environments indicating that this species is only located in the very top section of the SiO\textsubscript{2} layer. After NO treatment the nitrogen is always confined to the interface. If we define the interface as a region of $\pm$0.5 nm around the SiC/SiO\textsubscript{2} junction, we can use the DDF model to calculate the interface contribution relative to the normalised SiC signal as 0.14, 0.10 and 0.07 for 4, 6 and 9 keV, respectively. In agreement with this model, the associated Si-C-N environment varies slightly across the excitation energies. The observed Si-C-N environment stems from the passivation of dangling carbon bonds on the SiC side of the interface, which are one of the major active defects suggested to populate the SiC/SiO\textsubscript{2} interface.\cite{Gruber2018ElectricallyInterface} The NO nitridation of SiC/SiO\textsubscript{2} stacks is effective in reducing interface defects and improving the overall device behaviour as the nitrogen preferentially reacts at the interface to compensate these defects.\cite{Regoutz2018InterfaceAtmospheres,Berens2019NH3Reliability} In all previous SXPS experiments only a single chemical environment was reported in NO treated samples. However, peak fit analysis of the HAXPES spectra (see Figure \ref{fig:N_CL_comp2}) reveals that the peak has a clear asymmetry towards higher BE stemming from the presence of a second environment, which was most likely destroyed during sputtering in previous SXPS studies. The higher BE component has a binding energy close to Si-O-N environments. The feature behaves similarly to the main Si-C-N environment with its intensity being almost constant across excitation energies. Whilst the nitrogen can compensate Si-C dangling bonds on the carbide side of the interface resulting in Si-C-N environments, it also clearly helps to compensate defects on the oxide side of the interface by forming Si-O-N structures.\\

\section{Conclusion}
The present work shows how energy-dependent HAXPES can be applied to device-relevant multilayer structures to study elemental distributions and chemical environments across a multilayer system. Here, it was specifically used to study the effects of nitridation on the bulk layers as well as the interface of SiC/SiO\textsubscript{2} stacks. The non-destructive nature of HAXPES combined with careful peak fit analysis allows for the first time the exploration of previously undetected features providing more detailed information on how nitrogen compensates defects in these device structures. In the NO treated sample it is particularly important that HAXPES not only detects the Si-C-N states compensating Si-C dangling bonds on the carbide side of the interface, but also makes Si-O-N environments compensating oxide defects close to the interface observable. This further manifests the current status of NO as the industry standard to compensate interface defects. Important aspects of the complex nature of the incorporation of nitrogen in the SiO\textsubscript{2} layer for the NH\textsubscript{3} and NO+NH\textsubscript{3} treated samples are uncovered, providing crucial insights into the pros and cons of the application of such annealing atmospheres to device structures. Si-N and N-O environments, which could not be detected and identified previously, are found to be present within the oxide layer.  Modelling of the depth distribution function allows to quantify the information depth and signal contribution from different sections of the SiC/SiO\textsubscript{2} device stack. The results presented here increase our understanding of this critical interface and help to further inform the selection and optimisation of nitridation processes for SiC-based power electronics.

\ack
AR acknowledges the support from the Analytical Chemistry Trust Fund for her CAMS-UK Fellowship and from Imperial College London for her Imperial College Research Fellowship. NF acknowledges support from the Engineering and Physical Sciences Research Council (EP/L015277/1). This work was partly funded by the Austrian Research Promotion Agency (FFG, Project No. 863947). J.M.K. and J.L. acknowledge support from EPRSC under Grant No. EP/R002010/1. This work was carried out with the support of the Diamond Light Source, beamline I09 (proposal SI19885-1). The authors would like to thank Dave McCue, I09 beamline technician, for his support of the experiments. We would like to thank Joseph Woicik and Conan Weiland from the Materials Measurement Science Division at the National Institute of Standards and Technology and Alexander L. Shluger from University College London for fruitful discussions.

\section*{References}

\bibliographystyle{iopart-num}
\bibliography{references2}

\providecommand{\newblock}{}
\begin{thebibliography}{10}
\expandafter\ifx\csname url\endcsname\relax
  \def\url#1{{\tt #1}}\fi
\expandafter\ifx\csname urlprefix\endcsname\relax\def\urlprefix{URL }\fi
\providecommand{\eprint}[2][]{\url{#2}}

\bibitem{2019PowerGrid}
Bose B~K (ed) 2019 {\em {Power electronics in renewable energy systems and
  smart grid}\/} (Hoboken: John Wiley {\&} Sons) ISBN 9781119515661

\bibitem{Fiorenza2019CharacterizationReview}
Fiorenza P, Giannazzo F and Roccaforte F 2019 {\em energies\/} {\bf 12} 2310

\bibitem{Kimoto2014FundamentalsApplications}
Kimoto T and Cooper J~A 2014 {\em {Fundamentals of Silicon Carbide Technology:
  Growth, Characterization, Devices and Applications}\/} (IEEE Press and John
  Wiley {\&} Sons, Inc.)

\bibitem{Ferreira2019RoadmapEdition}
Ferreira B 2019 {\em {Roadmap for Wide Bandgap Power Semiconductors 2019
  Edition}\/} (IEEE)

\bibitem{Weitzel1996SiliconDevices}
Weitzel C~E, Palmour J~W, Carter C~H, Moore K, Nordquist K~J, Alien S, Thero C
  and Bhatnagar M 1996 {\em IEEE Transactions on Electron Devices\/} {\bf 43}
  1732--1741

\bibitem{Kimoto2015MaterialAnnealing}
Kimoto T 2015 {\em Japanese Journal of Applied Physics\/} {\bf 54} 040103

\bibitem{She2017ReviewAnd}
She X, Huang A~Q and Ozpineci B 2017 {\em IEEE Transactions on Industrial
  Electronics\/} {\bf 64} 8193--8205

\bibitem{Deak2007TheSiC/SiO2interface}
De{\'{a}}k P, Knaup J~M, Hornos T, Thill C, Gali A and Frauenheim T 2007 {\em
  Journal of Physics D: Applied Physics\/} {\bf 40} 6242--6253

\bibitem{Onneby1997SiliconInterface}
{\"{O}}nneby C and Pantano C~G 1997 {\em Journal of Vacuum Science {\&}
  Technology A: Vacuum, Surfaces, and Films\/} {\bf 15} 1597--1602

\bibitem{Kobayashi2003Interfacemn0001/m}
Kobayashi H, Sakurai T, Takahashi M and Nishioka Y 2003 {\em Physical Review
  B\/} {\bf 67} 115305

\bibitem{AminiMoghadam2016ActiveReview}
Amini~Moghadam H, Dimitrijev S, Han J and Haasmann D 2016 {\em Microelectronics
  Reliability\/} {\bf 60} 1--9

\bibitem{Gruber2018ImpactInterface}
Gruber G, Gspan C, Fisslthaler E, Dienstleder M, Pobegen G, Aichinger T,
  Meszaros R, Grogger W and Hadley P 2018 {\em Advanced Materials Interfaces\/}
   1800022

\bibitem{Pitthan2015SiCFilms}
Pitthan E, Gobbi A, Boudinov H and Stedile F 2015 {\em Journal of Electronic
  Materials\/} {\bf 44} 2823--2828

\bibitem{Baumvol1996ThermalHydrogen}
Baumvol I~J, Stedile F~C, Ganem J~J, Trimaille I and Rigo S 1996 {\em Journal
  of the Electrochemical Society\/} {\bf 143} 1426--1434

\bibitem{Li1999InvestigationSpectroscopy}
Li H~F, Dimitrijev S, Sweatman D, Harrison H~B, Tanner P and Feil B 1999 {\em
  Journal of Applied Physics\/} {\bf 86} 4316--4321

\bibitem{Jamet2001PhysicalSiC}
Jamet P and Dimitrijev S 2001 {\em Applied Physics Letters\/} {\bf 79} 323--325

\bibitem{Woerle2019Two-dimensionalInterface}
Woerle J, Johnson B~C, Bongiorno C, Yamasue K, Ferro G, Dutta D, Jung T~A, Sigg
  H, Cho Y, Grossner U and Camarda M 2019 {\em Physical Review Materials\/}
  {\bf 3} 84602

\bibitem{Isomura2019DistinguishingSpectroscopy}
Isomura N, Kutsuki K, Kataoka K, Watanabe Y and Kimoto Y 2019 {\em Journal of
  Synchrotron Radiation\/} {\bf 26} 462--466

\bibitem{Cottom2018RecombinationCalculations}
Cottom J, Gruber G, Pobegen G, Aichinger T and Shluger A~L 2018 {\em Journal of
  Applied Physics\/} {\bf 124}

\bibitem{Umeda2019ElectricallyOxidation}
Umeda T, Kagoyama Y, Tomita K, Abe Y, Sometani M, Okamoto M, Harada S and
  Hatakeyama T 2019 {\em Applied Physics Letters\/} {\bf 115} 0--5

\bibitem{Umeda2020CarbonInterface}
Umeda T, Kobayashi T, Sometani M, Yano H, Matsushita Y and Harada S 2020 {\em
  Applied Physics Letters\/} {\bf 116}

\bibitem{Hornetz1994ARXPSSurfaces}
Hornetz B, Michel H~J and Halbritter J 1994 {\em Journal of Materials
  Research\/} {\bf 9} 3088--3094

\bibitem{Zhu2011ChemicalSpectroscopy}
Zhu Q, Huang L, Li W, Li S and Wang D 2011 {\em Applied Physics Letters\/} {\bf
  99} 2--5

\bibitem{Woerle2017ElectronicARPES}
Woerle J, Bisti F, Husanu M~A, Strocov V~N, Schneider C~W, Sigg H, Gobrecht J,
  Grossner U and Camarda M 2017 {\em Applied Physics Letters\/} {\bf 110}
  132101

\bibitem{Regoutz2018InterfaceAtmospheres}
Regoutz A, Pobegen G and Aichinger T 2018 {\em Journal of Materials Chemistry
  C\/} {\bf 6} 12079--12085

\bibitem{Hamada2017AnalysisNO-POA}
Hamada K, Mikami A, Naruoka H and Yamabe K 2017 {\em e-Journal of Surface
  Science and Nanotechnology\/} {\bf 15} 109--114

\bibitem{Yamashita2019SpectroscopicInterface}
Yamashita Y, Nagata T, Chikyow T, Hasunuma R and Yamabe K 2019 {\em e-Journal
  of Surface Science and Nanotechnology\/} {\bf 17} 56--60

\bibitem{Chung2000EffectCarbide}
Chung G~Y, Tin C~C, Williams J~R, McDonald K, Di~Ventra M, Pantelides S~T,
  Feldman L~C and Weller R~A 2000 {\em Applied Physics Letters\/} {\bf 76}
  1713--1715

\bibitem{McDonald2003CharacterizationSiO2/4H-SiC}
McDonald K, Weller R~A, Pantelides S~T, Feldman L~C, Chung G~Y, Tin C~C and
  Williams J~R 2003 {\em Journal of Applied Physics\/} {\bf 93} 2719--2722

\bibitem{Regoutz2018ASystem}
Regoutz A, Mascheck M, Wiell T, Eriksson S~K, Liljenberg C, Tetzner K,
  Williamson B~A, Scanlon D~O and Palmgren P 2018 {\em Review of Scientific
  Instruments\/} {\bf 89}

\bibitem{Lee2018ASource}
Lee T~L and Duncan D~A 2018 {\em Synchrotron Radiation News\/} {\bf 31} 16--22

\bibitem{Shinotsuka2019CalculationsAlgorithm}
Shinotsuka H, Tanuma S, Powell C~J and Penn D~R 2019 {\em Surface and Interface
  Analysis\/} {\bf 51} 427--457

\bibitem{Solokha2018EffectiveMembranes}
Solokha V, Lee T~L, Wilson A, Hingerl K and Zegenhagen J 2018 {\em Journal of
  Electron Spectroscopy and Related Phenomena\/} {\bf 225} 28--35

\bibitem{Berens2019NH3Reliability}
Berens J, Pobegen G, Rescher G, Aichinger T and Grasser T 2019 {\em IEEE
  Transactions on Electron Devices\/} {\bf 66} 4692--4697

\bibitem{Gruber2018ElectricallyInterface}
Gruber G, Cottom J, Meszaros R, Koch M, Pobegen G, Aichinger T, Peters D and
  Hadley P 2018 {\em Journal of Applied Physics\/} {\bf 123}

\end{thebibliography}

\end{document}